\documentclass[aps,preprint]{revtex4}%
\usepackage{amsfonts}
\usepackage{amsmath}
\usepackage{amssymb}
\usepackage{graphicx}%
\begin{document}
\title{{\LARGE ON THE ROTATIONAL AND MACHIAN PROPERTIES OF THE UNIVERSE }}
\author{Marcelo Samuel Berman$^{1}$}
\affiliation{$^{1}$Instituto Albert Einstein / Latinamerica - Av. Candido Hartmann, 575 -
\ \# 17}
\affiliation{80730-440 - Curitiba - PR - Brazil email: msberman@institutoalberteinstein.org}
\date{(last version 11 July, 2008)}

\begin{abstract}
We find a Classical explanation on the origin of the Cosmological "constant"
term, as a rotating feature of the Universe. We give a picture on "creation"
of the Universe.

By analogy with the original Brans-Dicke relation $\frac{GM}{c^{2}R}\sim1$ \ ,
several other similar relations were obtained in a previous paper (Berman,
2007d), which we now extend, by relating the angular momentum \ $L$\ \ ,
\ \ absolute temperature \ $T$\ , and the cosmological "constant" \ $\Lambda
$\ \ \ , with \ $R$\ , where \ $R$\ , \ $M$\ \ \ stand for the radius
(scale-factor) and the mass of the causally related Universe. We show here
that these properties, which characterize Machian Universes, can be put in the
form of equalities, which can be derived, from the conjecture that the total
energy of the Universe is null, alike with the effective energy density, and
we therefore imply that \ \ $R$\ \ depends linearly with \ $M$\ \ ; the
angular momentum runs like \ $R^{2}$\ ; \ \ $T$\ \ \ , \ like \ $R^{-\frac
{1}{2}}$\ \ ; lambda, as \ $R^{-2}$\ \ so that the total entropy of the
Universe grows \ with \ $R^{\frac{3}{2}}$\ , and, then, also as \ $M^{\frac
{3}{2}}$\ \ .\ The "loss of information paradox", for black holes, is
dismantled, and, if the total entropy, in its mass dependence with
\ $M^{\frac{3}{2}}$, or with the scale-factor (Schwarzschild's radius),
\ $R^{\frac{3}{2}}$\ shall be the same for black holes and the Universe, we
hint that the Universe is very similar to a white-hole (Berman, 2007 b; 2007
a; 2007 c).

It must be obvious, from this paper, that each type of energy contribution, to
the total energy of the Universe, divided by \ $Mc^{2}$\ , yields\ a constant,
during all times, so that, for instance, if the present contribution of \ the
cosmological "constant"\ $\Lambda$\ \ , drives the present Universe, it also
must have driven alike, in all the lifespan of the Machian Universe: the
relative contributions of densities of each kind, towards the total density,
remain the same during all times. This fact is supported by the recent
discovery that the Universe has been accelerating since a long time ago.

Berman (2008; 2008a; 2008b) found similar hidden rotational states of the
Universe in Relativistic theories which obey Robertson-Walker's metric. 

\end{abstract}
\maketitle

\begin{center}
{\LARGE ON THE ROTATIONAL AND MACHIAN PROPERTIES OF THE UNIVERSE }

\bigskip

Marcelo Samuel Berman
\end{center}

\bigskip

\bigskip\bigskip{\large I. INTRODUCTION}

\bigskip Three recently published paper by \bigskip Berman (2007d, 2008a,
2008b), have presented the views of the present author on Machian Universes
and/or rotational hidden features of standard cosmology (see also, Berman,
2008). Here, we extend the Machian picture, we propose that the given picture,
as in the three papers cited above, points to a possible explanation of the
Classical origin of the cosmological "constant", as a centrifugal
acceleration, a possible asymmetric matter content explanation, etc.

In the beginning of the Universe, there was "nothing". We can say then, that
the initial total energy of the Universe was zero-valued. By assuming the
conservation of total energy, we may say that, up to now, the Universe is also zero-total-energy-valued.

The Universe has been \ $\Lambda$\ -- driven, for the most part of its
history. In this paper, we present the Machian Universe, in such a way, that
the above fact, is evident.\ 

\bigskip

One of the purposes of the present work, is to show, first, that if the
Machian Universe is depicted as having zero-total energy, not only it must
obey several "generalized" Brans-Dicke equalities, but, the resulting
cosmological model, has no initial singularity. Then, we shall see that we
must also include a rotational aspect of the Universe, which validate the
relation, valid for the angular speed, 

\bigskip

$\omega\propto R^{-1}$ \ \ \ \ \ \ \ \ \ \ \ \ \ \ \ \ \ \ \ \ \ \ \ ,

\bigskip

\bigskip where \ \ $R$\ \ \ is the radius of the causally related\ Universe,
which is proportional to the scale-factor of Robertson-Walker's metric, where
\ we may fix the scale-factor magnitude arbitrarily, without forgetting that
it represents a growing scale. This relation was found by Berman, in the three
above cited papers.

\bigskip

Barbour and Pfister(1995), listed several possible meanings generally
attributed to a Machian Universe, among others, as: a) the problem of motion
definition; b) the determination of inertial frames; c) the interactions
between masses; d) the generation of inertia; e) the induction of inertial
forces by means of accelerations, in analogy with electromagnetism; f) the
requirement that the metric tensor should be determined by matter; or by
matter and geometric degrees of freedom for gravity; g) the cosmic derivation
of inertial mass by means of the ADM Hamiltonian; h) the generalization of
Special Relativity when gravitation is present; i) the requirement of general
covariance of physical laws; j) the existence, or not, of boundary conditions
for the Universe; or the existence or inexistence of matter-free
singularity-free solutions in Cosmology; k) the requirement of physical
causality in Nature; l) the prediction of dragging effects due to the
distribution of masses; m) the antidote to G\"{o}del-type Universes; n) the
explanation of the origin of Brans-Dicke relation; o) the selection among
possible cosmological models; p) the non-existence of an absolute dynamical
theory; etc.

\bigskip

\bigskip\bigskip{\large II. A THEORY FOR MACHIAN UNIVERSES}

We now shall propose that Mach's Principle, means a zero-total energy
Universe(Feynman, 1962-3). Berman(2006; 2006a), has shown this meaning of
Mach's Principle without considering a rotating Universe. Berman has included
the spin of the Universe, \ and replaced Brans-Dicke traditional relation,
\ $\frac{GM}{c^{2}R}\sim1$\ \ , with three different relations, which\ we call
the Brans-Dicke relations for gravitation, for the cosmological "constant"\ ,
and for the spin of the Universe (Berman, 2007a; 2007c; 2007d). We now extend
the notion of Machian Universe, by including radiation.

\bigskip

We shall consider a "large" sphere, with mass \ $M$\ \ , radius \ $R$\ \ ,
spin \ $L$\ \ , and endowed with a cosmological term \ $\Lambda$\ , which
causes the existence\ \ of an energy density \ \ \ \ $\frac{\Lambda}{\kappa}%
$\ \ \ , where \ $\kappa=\frac{8\pi G}{c^{2}}$\ . We now calculate the total
energy\ \ $E$\ \ of this distribution:

\bigskip

$E=E_{i}+E_{g}+E_{L}+E_{\Lambda}+E_{R}$\ \ \ \ \ \ \ \ \ \ \ \ \ , \ \ \ \ \ \ \ \ \ \ \ \ \ \ \ \ \ \ \ \ \ \ \ \ \ \ \ \ \ \ \ \ \ (1)

\bigskip

where \ \ $E_{i}=Mc^{2}$\ \ , stands for the inertial (Special
Relativistic)\ \ energy; \ $E_{g}\cong-\frac{GM^{2}}{R}$\ \ \ \ \ (the
Newtonian gravitational potential self-energy); \ \ $E_{L}\cong\frac{L^{2}%
}{MR^{2}}$\ \ the Newtonian rotational energy; $\ E_{\Lambda}\cong
\frac{\Lambda R^{3}}{6G}\ $\ (the cosmological "constant" energy contained
within the sphere), and \ $E_{R}=aT^{4}$\ \ , where \ $a$\ \ is a constant,
and \ $T$\ \ \ stands for absolute temperature, while \ $E_{R}$\ \ represents
radiational energy.\ \ 

\bigskip

If we impose that the total energy is equal to zero, i.e., $E=0$\ \ , and
\ $\dot{E}=0$\ \ ( the total energy is null, and time-invariant\ ), we obtain
from (1):

$\frac{GM}{c^{2}R}-\frac{L^{2}}{M^{2}c^{2}R^{2}}-\frac{\Lambda R^{3}}%
{6GMc^{2}}-\frac{E_{R}}{Mc^{2}}\cong1$\ \ \ \ \ \ \ \ \ \ \ \ . \ \ \ \ \ \ \ \ \ \ \ \ \ \ \ \ \ \ \ \ \ \ \ \ \ \ \ \ \ (2)

\bigskip

\bigskip As relation (2) above should be valid for the whole Universe, and not
only for \ a specific instant of time, in the life of the Universe, and if
this is not a coincidental relation, we can solve this equation by imposing that:

\bigskip

$\frac{GM}{c^{2}R}=\gamma_{G}\sim1$ \ \ \ \ \ \ \ \ \ \ \ \ \ \ \ \ \ \ \ \ , \ \ \ \ \ \ \ \ \ \ \ \ \ \ \ \ \ \ \ \ \ \ \ \ \ \ \ \ \ \ \ \ \ \ \ \ \ \ \ \ \ \ \ \ \ \ \ \ \ (3)

\bigskip

$\frac{L}{McR}=\gamma_{L}$%
\ \ \ \ \ \ \ \ \ \ \ \ \ \ \ \ \ \ \ \ \ \ \ \ \ \ \ ; \ \ \ \ \ \ \ \ \ \ \ \ \ \ \ \ \ \ \ \ \ \ \ \ \ \ \ \ \ \ \ \ \ \ \ \ \ \ \ \ \ \ \ \ \ \ \ \ \ \ (4)

\bigskip

\bigskip$\frac{E_{R}}{Mc^{2}}=\frac{\rho_{R}R^{3}}{Mc^{2}}=\frac{aT^{4}R^{3}%
}{Mc^{2}}=\gamma_{R}\sim10^{-3}$ \ \ \ \ \ \ \ \ \ \ , \ \ \ \ \ \ \ \ \ \ \ \ \ \ \ \ \ \ \ \ \ \ \ \ \ \ \ \ \ \ (4a)

and,

\bigskip

$\frac{\Lambda R^{3}}{6GMc^{2}}=\gamma_{\Lambda}$%
\ \ \ \ \ \ \ \ \ \ \ \ \ \ \ \ \ \ \ \ \ \ \ \ \ \ , \ \ \ \ \ \ \ \ \ \ \ \ \ \ \ \ \ \ \ \ \ \ \ \ \ \ \ \ \ \ \ \ \ \ \ \ \ \ \ \ \ \ \ \ \ \ \ \ \ (5)

\bigskip

subject to the condition,

\bigskip

$\gamma_{G}-\gamma_{L}^{2}-\gamma_{\Lambda}-\gamma_{R}\cong1$%
\ \ \ \ \ \ \ \ \ \ \ \ \ \ , \ \ \ \ \ \ \ \ \ \ \ \ \ \ \ \ \ \ \ \ \ \ \ \ \ \ \ \ \ \ \ \ \ \ \ \ \ \ \ \ \ \ \ (6)

\bigskip

where the \ $\gamma^{\prime}s$\ \ are constants.

\bigskip

It must be remarked, that our proposed law (3), is a radical departure from
the original Brans-Dicke (Brans and Dicke, 1961) relation, which was an
approximate one, while our present hypothesis implies that \ \ $R\propto
M$\ \ . With the present hypothesis, one can show, that independently of the
particular gravitational theory taken as valid, the energy density of the
Universe obeys a \ \ $R^{-2}$\ \ dependence (see Berman, 2006; 2006a; Berman
and Marinho, 2001).

\bigskip

More than that, we ensure, with the conditions (3)(4)(4a) and (5), that each
different type of relative energy contribution, into the total energy, remains
the same, during all periods of time, when compared with \ \ $Mc^{2}$\ .\ 

We have now the following generalized Brans-Dicke relations, for gravitation,
spin, radiation and cosmological "constant":

\bigskip

$\frac{GM}{c^{2}R}=\gamma_{G}\sim1$ \ \ \ \ \ \ \ \ \ \ \ \ \ \ \ \ \ \ \ \ , \ \ \ \ \ \ \ \ \ \ \ \ \ \ \ \ \ \ \ \ \ \ \ \ \ \ \ \ \ \ \ \ \ \ \ \ \ \ \ \ \ \ \ \ \ \ \ \ \ (3)

\bigskip

$\frac{GL}{c^{3}R^{2}}=\gamma_{G}$ $.$ $\gamma_{L}$
\ \ \ \ \ \ \ \ \ \ \ \ \ \ \ \ \ \ \ \ \ , \ \ \ \ \ \ \ \ \ \ \ \ \ \ \ \ \ \ \ \ \ \ \ \ \ \ \ \ \ \ \ \ \ \ \ \ \ \ \ \ \ \ \ \ \ \ \ (7)

\bigskip

\bigskip$\frac{aGT^{4}R^{2}}{c^{4}}=\gamma_{R}$ $.$ $\gamma_{G}\sim10^{-3}$ \ \ \ \ \ \ \ \ \ \ ,\ \ \ \ \ \ \ \ \ \ \ \ \ \ \ \ \ \ \ \ \ \ \ \ \ \ \ \ \ \ \ \ \ \ \ \ \ \ \ \ \ \ (7a)

and,

\bigskip

$\frac{\Lambda R^{2}}{6c^{4}}=\gamma_{\Lambda}$ $.$ $\gamma_{G}\sim1$
\ \ \ \ \ \ \ \ \ \ \ \ \ \ . \ \ \ \ \ \ \ \ \ \ \ \ \ \ \ \ \ \ \ \ \ \ \ \ \ \ \ \ \ \ \ \ \ \ \ \ \ \ \ \ \ \ \ \ \ \ \ (8)

\bigskip

The reader should note that we have termed \ $\Lambda$\ \ as a "constant", but
it is clear from the above, that in an expanding Universe, \ $\Lambda\propto
R^{-2}$\ \ , so that \ $\Lambda$\ is a variable term.\ We also notice that
\ $R\propto M$\ \ \ , and \ $L\propto R^{2}$\ , and \ $R\propto T^{-2}$\ \ ,
so that , we should have,\ \ $\rho_{R}\propto R^{-2}$\ \ \ . The \ $T^{-2}%
$\ \ dependence on \ $R$\ \ , was dealt, earlier, for non-relativistic
decoupled\ massive species (Kolb and Turner, 1990).

\bigskip

It is clear from our previous hypotheses, that all the energy densities vary
with \ $R^{-2}$\ . This can be checked one by one. For instance, from the
definition of the inertial energy density,

\bigskip

$\rho_{i}=\frac{M}{V}$\ \ \ \ \ \ \ \ \ , \ \ \ \ \ \ \ \ \ \ \ \ \ \ \ \ \ \ \ \ \ \ \ \ \ \ \ \ \ \ \ \ \ \ \ \ \ \ \ \ \ \ \ \ \ \ \ \ \ \ \ \ \ \ \ \ \ \ \ \ \ \ \ \ (9)

\bigskip

while, \ \ 

\bigskip

$V=\alpha R^{3}$\ \ \ \ \ \ ,\ \ \ \ \ ( $\alpha=$\ constant\ \ ) \ \ \ \ \ \ \ \ \ \ \ \ \ \ \ \ \ \ \ \ \ \ \ \ \ \ \ \ \ \ \ \ \ \ \ \ (10)

\bigskip

where \ $\rho_{i}$\ \ and \ \ $V$\ \ stand for the inertial (or, matter)
energy density and tridimensional volume, we find:

\bigskip

$\rho_{i}=\left[  \frac{\gamma_{G}}{G\text{ }\alpha}\right]  R^{-2}%
$\ \ \ \ \ \ \ \ \ . \ \ \ \ \ \ \ \ \ \ \ \ \ \ \ \ \ \ \ \ \ \ \ \ \ \ \ \ \ \ \ \ \ \ \ \ \ \ \ \ \ \ \ \ \ \ \ \ \ \ \ \ \ \ (11)

\bigskip

\bigskip\bigskip B.D. approximate relation for spin, has been derived,
earlier, on a heuristic procedure, which consists on the simple hypothesis
that \ $L$\ \ should obey a similar relation as $M$\ \ (Sabbata and Sivaram,
1994).\ The first authors to propose the above \ $R^{-2}$\ \ dependence for
$\Lambda$\ \ \ were Chen and Wu (1990), under the hypothesis that $\Lambda
$\ \ should not depend on Planck's constant, because the cosmological
"constant"\ \ is the Classical Physics response to otherwise Quantum effects
that originated the initial energy of the vacuum. Berman, as well as Berman
and Som, have examined, along with other authors, a time dependence for
$\Lambda$ \ \ (see for example, Berman, 1991; 1991a).

\bigskip

If we apply the above relation, for Planck's and the present Universe, we find:

\bigskip

$\frac{\rho_{i}}{\rho_{Pl}}=\left[  \frac{R}{R_{Pl}}\right]  ^{-2}$
\ \ \ \ \ \ \ \ \ \ \ . \ \ \ \ \ \ \ \ \ \ \ \ \ \ \ \ \ \ \ \ \ \ \ \ \ \ \ \ \ \ \ \ \ \ \ \ \ \ \ \ \ \ \ \ \ \ \ \ \ \ \ (12)

\bigskip

If we substitute the known values for Planck's quantities, while we take for
the present Universe, \ $R\cong10^{28}$\ cm, we find a reasonable result for
the present matter energy density. This shows that our result (relation 11),
has to be given credit. For the energy density of the \ $\Lambda$\ - term, we
shall have \ $\rho_{\Lambda}=\frac{\Lambda}{\kappa}\propto R^{-2}$\ \ , and
also for the rotational energy, we have \ $\rho_{L}=\frac{L^{2}}{c^{2}%
M^{2}R^{4}}\propto R^{-2}$\ \ ; we have also shown above that the radiation
energy density is also proportional to \ $R^{-2}$\ . We conclude that all
forms of energy densities are \ $R^{-2}$\ \ - dependent, provided that the
generalized Brans-Dicke relations, for gravitation, spin, cosmological
"constant" and radiation, obey the conditions \ (3), (7), (7a) and (8)\ above.\ \ \ \ \ \ 

\bigskip

\bigskip We point out that, what we call the total energy density of the
Universe, is the sum of all the positive energy densities for everything
except the self-gravitational energy density, which is negative, and makes the
effective total energy density of the Universe, zero-valued, and
time-independent. This is how the Machian Universe escapes from the accusation
of keeping the initial singularity of the Universe: we are sure, because of
the time-invariance of the null results for total energy and effective energy
density, that,

\bigskip

$\lim\limits_{R\rightarrow0}\rho=\lim\limits_{R\rightarrow0}E=0$%
\ \ \ \ \ \ \ \ \ \ \ . \ \ \ \ \ \ \ \ \ \ \ \ \ \ \ \ \ \ \ \ \ \ \ \ \ \ \ \ \ \ \ \ \ \ \ \ \ \ \ \ \ \ \ \ \ \ \ \ (11b)

We show now that the entropy of the Machian Universe, as described above, is
not constant, but increases with time. On defining, \ \ \ 

\bigskip

$S=sR^{3}$ \ \ \ \ \ \ \ \ \ \ \ \ \ ,

\bigskip

where, \ \ \ $S$\ \ and \ $s$\ \ \ represent total entropy, and entropy
density, respectively for the Universe. According to the above formulae, we have:

\bigskip

$S\propto sR^{3}=\frac{\rho_{R}}{T}R^{3}\propto T^{3}R^{3}\propto R^{\frac
{3}{2}}\propto M^{\frac{3}{2}}$ \ \ \ \ \ \ \ \ , \ \ \ \ \ \ \ \ \ \ \ \ \ \ \ \ \ \ \ \ \ \ \ \ \ \ \ \ \ \ \ (12a)

\bigskip

taken care\ \ of the relations given earlier, i.e., \ $R\propto T^{-2}$ \ ,
and (3) above\ \ .

\bigskip

We have thus shown that \ $S$\ \ increases with time, likewise the scale
factor to the power $\frac{3}{2}$\ .[This Machian property stands opposite to
the usual assumptions in theoretical studies that, by considering \ $S=$
\ constant\ , found \ \ $R\propto T^{-1}$\ \ \ , which as we have shown, is
incorrect in the Machian picture].\ 

\bigskip

\bigskip For a black hole, \ the usual black body entropy formula (Sears and
Salinger, 1975), yields similar equations, when we substitute the scale factor
by Schwarzschild's \ radius.

This solves the "loss of information paradox" , for black holes, because this
body fills a \ bounded region of space, limited by the horizon. As the
Universe is isotropic and homogeneous, each bounded region of it, must share
the growth in entropy of the whole Universe, because \ the mass of black hole,
being part of the Universe, increases likewise, with the mass of the
Universe.\ \ \bigskip However, for clothed singularities, we have locally, a
fixed event horizon radius, the Schwarzschild's one. Then, the entropy is
locally constant; so, the only way to increase it, is by accretion, whereby,
in the local scenario, there is an exchange between the entropies of the local
neighborhood and the black hole, but the total local entropy is constant:
there is no total local loss of information (Berman, 2007). If the black-hole
singularity is naked, and because the Universe is homogeneous and isotropic
while the black-hole symmetries turn the same amount of entropy lost by the
collapsing body, retrieved by the exterior of the black-hole, and there is no
"loss of information".

It should be remembered that the origin of Planck's quantities, say, for
length, time, density and mass, were obtained by means of
dimensional\ \ combinations among the constants for macrophysics ($G$ for
gravitation and $c$ for electromagnetism) and for Quantum Physics (Planck's
constant $\frac{h}{2\pi}$). Analogously, if we would demand a dimensionally
correct Planck's spin, obviously we would find,

\bigskip

$L_{Pl}=\frac{h}{2\pi}$ \ \ \ \ \ \ \ \ \ \ \ \ \ \ . \ \ \ \ \ \ \ \ \ \ \ \ \ \ \ \ \ \ \ \ \ \ \ \ \ \ \ \ \ \ \ \ \ \ \ \ \ \ \ \ \ \ \ \ \ \ \ \ \ \ \ \ \ \ \ \ \ \ \ \ \ \ \ \ (13)

\bigskip

From Brans-Dicke relation for spin, we now can obtain the present angular
momentum of the Universe,

$\bigskip$

$L=L_{Pl}\left[  \frac{R}{R_{Pl}}\right]  ^{2}\cong10^{120}\left(  \frac
{h}{2\pi}\right)  =10^{93}$ $\ \ g$ $cm^{2}$ $s^{-1}$ \ \ \ \ \ \ \ . \ \ \ \ \ \ \ \ \ \ \ \ \ \ \ \ \ (14)

\bigskip

This estimate was also made by Sabbata and Sivaram(1994), based on heuristic
considerations(see also Sabbata and Gasperini, 1979).

\bigskip

If we employ, for the cosmological "constant" Planck's value, $\Lambda_{Pl}$ ,

\bigskip

$\Lambda_{Pl}\cong R_{Pl}^{-2}$ \ \ \ \ \ \ \ \ , \ \ \ \ \ \ \ \ \ \ \ \ \ \ \ \ \ \ \ \ \ \ \ \ \ \ \ \ \ \ \ \ \ \ \ \ \ \ \ \ \ \ \ \ \ \ \ \ \ \ \ \ \ \ \ \ \ \ \ \ \ \ \ \ \ (15)

\bigskip

then, we shall find, in close agreement with the present value estimate for
\ $\Lambda$\ (as found by recent supernovae observations), by means of the
third Brans-Dicke relation:

\bigskip

$\Lambda=\Lambda_{Pl}\left[  \frac{R_{Pl}}{R}\right]  ^{-2}$ \ \ \ \ . \ \ \ \ \ \ \ \ \ \ \ \ \ \ \ \ \ \ \ \ \ \ \ \ \ \ \ \ \ \ \ \ \ \ \ \ \ \ \ \ \ \ \ \ \ \ \ \ \ \ \ \ \ \ \ \ \ \ \ (16)

\bigskip

This is, however, the first time that the above results are obtained by means
of the zero-total energy hypothesis for the Universe. This is why we attribute
this hypothesis to a Machian Universe; indeed, we believe that we can identify
Mach's Principle, with this hypothesis. Then, the entropy of the Universe
grows with increasing time, because it is expanding, and so does the entropy
of any black hole, in the big global picture, but, locally, it remains
constant, for a clothed singularity, because \ $R_{S}$\ \ is constant.\ 

\bigskip

We remark that, as we showed earlier, $\ S$\ \ is proportional to
\ \ $M^{\frac{3}{2}}$\ \ \ , either for the Universe or the immersed\ \ black holes.

\bigskip

Berman (2007b) has depicted the Universe, as a "white" black hole; here, we
have shown, that both entropies coincide in the same \ $R^{\frac{3}{2}}%
$\ \ \bigskip-- \ dependence.

\bigskip From the Machian model of the Universe, as described through the
zero-total energy principle, we may show that the so-called initial
singularity problem, in Cosmology, is not a problem at all, for the
singularity is not essential: in fact, the limit of \ $\rho$\ \ and
\ \ $E$\ \ \ when \ $R\rightarrow0$\ \ , is not infinite, but zero (according
to the zero-total energy condition and relation (11b)).

\bigskip

\bigskip\bigskip{\large III. THE CLASSICAL ORIGIN FOR LAMBDA}

\bigskip If the Universe has attached a "spin", \ \ \ $L$\ \ , it has also an
angular speed. Consider the Classical definition, 

\bigskip

$L=MR^{2}\omega$ \ \ \ \ \ \ \ \ \ \ \ \ \ \ \ \ . \ \ \ \ \ \ \ \ \ \ \ \ \ \ \ \ \ \ \ \ \ \ \ \ \ \ \ \ \ \ \ \ \ \ \ \ \ \ \ \ \ \ \ \ \ \ \ \ \ \ \ \ \ \ \ \ \ \ \ (17)

\bigskip

From relation (3), we find, \ \ 

\bigskip

$L=\frac{c^{2}\gamma_{G}}{G}R^{3}\omega$\ \ \ \ \ \ \ \ \ \ \ \ \ \ . \ \ \ \ \ \ \ \ \ \ \ \ \ \ \ \ \ \ \ \ \ \ \ \ \ \ \ \ \ \ \ \ \ \ \ \ \ \ \ \ \ \ \ \ \ \ \ \ \ \ \ \ \ \ \ \ \ \ \ (18)

\bigskip

From (14), we know that \ \ $L\propto R^{2}$\ \ , so that, in (18), we have,

\bigskip

$\omega=\frac{\beta}{R}$\ \ \ \ \ \ \ \ \ \ \ \ \ \ \ \ \ \ \ \ . \ \ \ \ \ \ \ \ \ \ \ \ \ \ \ \ \ \ \ \ \ \ \ \ \ \ \ \ \ \ \ \ \ \ \ \ \ \ \ \ \ \ \ \ \ \ \ \ \ \ \ \ \ \ \ \ \ \ \ \ \ \ \ (19)

\bigskip

From physical considerations, because \ $\beta$\ \ \ is a speed, and because
of relation \ (14) , we must impose that \ $\beta\leq c$\ \ \ . \ But from
(13), we can see that the Machian angular speed should be, 

\bigskip

$\omega=\frac{c}{R}$\ \ \ \ \ \ \ \ \ \ \ \ \ \ \ \ \ \ \ \ , \ \ \ \ \ \ \ \ \ \ \ \ \ \ \ \ \ \ \ \ \ \ \ \ \ \ \ \ \ \ \ \ \ \ \ \ \ \ \ \ \ \ \ \ \ \ \ \ \ \ \ \ \ \ \ \ \ \ \ \ \ \ \ (20)

\bigskip

so that all the above relations be compatible with their values chosen for
Planck's Universe. It was the above formula, that prompted Berman to include a
centrifugal acceleration for the Universe, namely, 

\bigskip

$a_{cf}=\omega^{2}R$ \ \ \ \ \ \ \ \ \ \ \ \ \ \ \ \ \ . \ \ \ \ \ \ \ \ \ \ \ \ \ \ \ \ \ \ \ \ \ \ \ \ \ \ \ \ \ \ \ \ \ \ \ \ \ \ \ \ \ \ \ \ \ \ \ \ \ \ \ \ \ \ \ \ \ \ \ \ (21)

\bigskip

This acceleration accounts for the Pioneer anomaly (Berman, 2007d). It has
been recently announced in the media, that NASA researchers have found that
this acceleration is not a particularity affecting only Pioneer space-probes,
but is associated with non-closed orbits. Our model fits these findings,
because in closed orbits around the Sun, cosmological phenomena do not
manifest themselves, because of small distances that are involved; remember
that open-orbits extend to infinity.

\bigskip

When one considers Raychaudhuri's equation, for a perfect fluid, we find,
(Raychaudhuri, 1979)

\bigskip

$\frac{\ddot{R}}{R}=-\frac{4\pi G}{3}\left(  \rho+3p-\frac{\Lambda}{4\pi
G}\right)  $ \ \ \ \ \ \ \ \ \ \ \ \ \ \ . \ \ \ \ \ \ \ \ \ \ \ \ \ \ \ \ \ \ \ \ \ \ \ \ \ \ \ \ \ \ \ \ \ \ \ \ \ \ \ \ \ (22)

{\large  }

\bigskip We see that there are three "accelerations", the last one given by,

\bigskip$a_{\Lambda}=\frac{\Lambda}{3}R$ \ \ \ \ \ \ \ \ \ \ \ \ \ \ \ \ \ \ . \ \ \ \ \ \ \ \ \ \ \ \ \ \ \ \ \ \ \ \ \ \ \ \ \ \ \ \ \ \ \ \ \ \ \ \ \ \ \ \ \ \ \ \ \ \ \ \ \ \ \ \ \ \ \ \ \ \ \ \ \ \ (23)

\ \ \ 

\bigskip The \ $R$ \ -- dependence\ of (23) and\ \ (21) is strikingly\ the
same. We suggest that this feature gives a possible Classical origin for
dark-energy, which has been represented by the Cosmological "constant", and we
associate with the rotational state of the Universe. This surmounts the
arguments against the existence of a lambda-term, because it would stand out
of Classical Physics, and be included as a Quantum-originated term.

\bigskip

\bigskip{\large IV. ORIGIN OF BARION ASYMMETRY IN THE UNIVERSE}

\bigskip

Let us now consider the origin of "creation": a fluctuation of the vacuum,
would result in two masses, \  $+\Delta M$ \ \ , and, \ \  $-\Delta M$ \ \ .
Conservation of mass is guaranteed. The positive mass, includes an angular
momentum \ \ $-\frac{h}{2\pi}$\ \ , while the negative mass has a spin in the
opposite direction. Conservation of angular momentum is thus supplied.
Consider that each mass above generates one single Universe: the first extends
from \ \ $t=0$\ \ \ onwards, with increasing positive time-coordinate; the
second mass, which is negative, extends from \ $t=0$\ \ , backwards, for
negative decreasing\ \ times. This is coherent with Robertson-Walker's metric,
which does not distinguish between positive or negative time-coordinates as preferential.

\bigskip

Consider now that both Universes are not "tied". Our Universe, is the one with
positive mass. That is why barions are positive here. The rest of the story,
is given by the Machian or General Relativistic Theories, as considered in
Berman(2007d, 2008, 2008a, 2008b). Our zero-total-energy, is still available.
Conservation of energy is also guaranteed.

\bigskip{\large V. DISCUSSION ABOUT THE PRESENT MODEL }

\bigskip We may argue that (1) it would be unclear who should measure the
energy of the Universe, from the "outside"; (2) it would be unclear whether we
may use Newtonian expressions for the calculations; (3) it would be
mathematically impossible to derive several generalized Brans-Dicke
equalities, from a single equation describing the energy \ $E$\ ; \ (4) the
local energy-momentum conservation, described by the covariant divergence of
the energy-momentum tensor, would be no more valid, and therefore, the model
is inconsistent; (5) the large angular-momentum of the Universe, is not
astronomically confirmed; (6) this paper does not obey any viable theory of
Gravity, and it does not supply new results about the Universe; (7) the
Brans-Dicke relation is numerically verified for the present Universe, but the
generalized counterpart, which is an equality, is obviously also verified, so
that, nothing new has been provided, and, the coincidence has a lot of
uncertainty; (8) what Berman is doing, is just an exercise in dimensional
analysis, like has been earlier done for instance, by Dirac and Eddington; (9)
this theory is heuristic, and, thus, not necessarily scientific; it puts on
the same footing special relativistic terms (the rest-mass energy) Newtonian
terms (the Newton potential), and General Relativistic terms (the Cosmological
"constant"), by adding them in equation (1); (10) we consider black-body
radiation inside a volume of size \ $R$\ \ , with a typical photon's
wavelength proportional to \ $R$\ \ , and the total number of photons is
therefore a constant (black-body radiation implies that the volume is full of
photons). Each photon has energy \ $kT$\ \ , or so, which is also inversely
proportional to its wavelength. Then, \ $T\propto R^{-1}$\ \ , and this has
nothing to do with the Universe being Machian or not; relation (4a) implies
that \ $R$\ \ is constant; so with \ $T$\ \ , \ \ $S$\ \ , $\Lambda$%
\bigskip\ \ \ and \ \ $M$\ \ \ (unless \ $G$\ \ and/or \ \ $c$\ \ \ are
variable). So, Berman's model\ is not evolutionary, and also does not resemble
real. (11) Newton's potential \ \ $\frac{GM}{c^{2}R}$\ \ of order 1, as in
equation (3), is a non-sense, because the Newton gravitational self-energy can
only be defined in that way, when the Newtonian gravitational approximation is
valid, which in turn, requires \ $|\frac{GM}{c^{2}R}|<<1$\ \ , in contrast
with (3).\ \

However, we answer those "cons", with the following "pros": (A) allegations
about the energy of the Universe, and, precisely, about its zero-value, can be
traced to Feynman (Feynman, 1962-3), Rosen (Rosen, 1994-95), Cooperstock and
Israelit (1995), Hawking (2001) and many others. Berman has derived this from
Robertson-Walker's metric, so that it is a valid result in Relativistic
Cosmology, for any tri-curvature value (Berman, 2006, 2006a). The existence of
a "spectator" is a philosophical question, rather than a scientific one; (B)
Machian properties have been proposed in different gravity theories, so there
is no one single theory that owns such attributes (remember the origin of
Brans-Dicke theory); (C) the several generalized Brans-Dicke equalities,
derived here from the energy equation, are just, the most simple set of
solutions for the \ $E=\dot{E}=0$\ \ equation; (D) the mentioned solutions,
have very interesting properties: for instance, the relative contributions of
each type of energy towards the total amount, is time-independent. This fact
is coherent with the recently proclaimed and experimentally observed result
that the Universe has been \ \ lambda-dominated since long ago; (E)\ we never
told that "Machian" conditions only can mean "general relativistic" ones; (F)
you can not blame our paper for the fact that the angular momentum is high for
the present Universe, because we have derived from this result, that the
amount of angular velocity in the present Universe is small and it is
undetectable with present technological tools; (G) our framework is
relativistic, in the low Newtonian limit, but this could be called, also, a
Sciama gravitational theory (Sciama, 1953); (H) we can extend all forms of
energy densities towards Planck's time, by going back from the present: no
inconsistency with Planck's energy density would be found; (I) because we can
not suppose on the first stance, that the entropy is constant, we now answer
the final objection (\# 10): the argument presented against the model is
dependent on the hypothesis that \ $RT=$\ \ constant, and thus \ \ $S=$%
\ \ constant\ . However, we have shown that, if \ \ $\dot{R}\neq0$\ \ , and
\ \ $\dot{E}=E=$ \ $\dot{G}=\dot{c}=0$\ \ \ , \ we do NOT have such constancy
for \ $RT$\ \ , but, instead, we have \ $RT^{2}=$\ \ constant\ . In this case,
the total entropy grows with \ $R^{3/2}$\ \ , while, according to the
complaint, it would be constant. It remains open, the possibility, to be
featured in a new paper, of a static model of the Universe, with time-varying
\ $G$\ \ \ and \ \ $c$\ \ , the sort of thing that I had not dreamed about
before reading the complaint in \# 10; (J)\ the fact that we mix rest-mass
energy, with a cosmological constant and a newtonian potential, rests on the
possibility of defining a newtonian cosmology with lambda (see for instance
D'Inverno, 1992); the newtonian potential\ \ can be associated with special
relativity, because, in a given point of space, General Relativity is locally
Special relativistic, and Newtonian - as far as standard cosmology applies
(Barrow, 1988); the Brans-Dicke relation, shows that the Universe obeys the
challenged relation \ \ \ $\frac{GM}{c^{2}R}=\gamma_{G}$\ \ , with
\ $\gamma_{G}\sim1$\ (if we remove the last condition, our theory remains
basically intact).\ \ \ \ 

\bigskip

We refer to the extremely important book by Sabbata and Sivaram(1994), where
there are clues about the rotation of the Universe. The astronomical Pioneer
anomaly, and the astrophysical \ laws, like Blackett formula, which relates
spin and magnetic field, \ and Wesson's one, relating spin with mass, are
discussed by us in another paper (Berman, 2007 d).

\bigskip

{\large \bigskip IV. \bigskip CONCLUSION}

\bigskip We hinted that a Machian Universe could be understood by a zero-total
energy model. Several properties of the model were derived, as pertaining to
the most simple solution for the energy equation. Whether the present model is
to be accepted or not, we point out that, as far as present experimental
evidence is concerned, it is acceptable.

\bigskip

The inconvenient hypothesis that \ \ $S=$\ \ constant, as in standard
Cosmology, has now been substituted by an ever-increasing entropy, as long as
\ \ $\dot{R}>0$\ \ . This may be the most important result of our paper, other
important conclusions being those about the Classical origin of the
lambda-term and the picture of \ the non-singular "creation".\bigskip

\bigskip{\Large Acknowledgements}

\bigskip

The author gratefully thanks his intellectual mentors, Fernando de Mello
Gomide and M. M. Som, his colleagues Nelson Suga, Marcelo F. Guimar\~{a}es,
Antonio F. da F. Teixeira, and Mauro Tonasse; I am also grateful for the
encouragement by Albert, Paula, and Geni, and messages by Dimi Chakalov.
Arguments against the present model, were here cleared, and we thank those who
helped with critical opinions on earlier versions of this paper.

\bigskip

{\Large References}

\bigskip

Barbour, J.; Pfister, H. (eds)(1995) - \textit{Mach's Principle - from
Newton's Bucket to Quantum Gravity} \ Birkh\"{a}user, Boston.

Barrow, J.D. (1988) - \textit{The Inflationary Universe, }in
\textit{Interactions and Structures in Nuclei}, pp 135-150 (eds) R Blin Stoyle
and W D Hamilton, Adam Hilger, Bristol.

Berman,M.S. (1991) - GRG \textbf{23}, 465.

Berman,M.S. (1991a) - Physical Review \textbf{D43}, 1075.

\bigskip Berman,M.S. (2006) - \textit{Energy of Black-Holes and Hawking's
Universe \ }in \textit{Trends in Black-Hole Research, }Chapter 5\textit{.}
Edited by Paul Kreitler, Nova Science, New York.

Berman,M.S. (2006 a) - \textit{Energy, Brief History of Black-Holes, and
Hawking's Universe }in \textit{New Developments in Black-Hole Research},
Chapter 5\textit{.} Edited by Paul Kreitler, Nova Science, New York.

Berman,M.S. (2006b) - See Los Alamos Archives,
http://arxiv.org/abs/physics/0611007 .

Berman,M.S. (2007 a) - \textit{Introduction to General Relativity and the
Cosmological Constant Problem}, Nova Science, New York.

Berman,M.S. (2007 b) - \textit{Is the Universe a White-Hole?}, Astrophysics
and Space Science, \textbf{311} ,359-361.

Berman,M.S. (2007 c) - \textit{Introduction to General Relativistic and
Scalar-Tensor Cosmologies}, Nova Science, New York.

\bigskip Berman,M.S. (2007 d) - \textit{The Pioneer's Anomaly and a Machian
Universe}, Astrophysics and Space Science, \textbf{312,} 275.

Berman,M.S. (2008) - \textit{A Primer in Black-Holes, Mach's Principle, and
Gravitational Energy }, Nova Science, New York.

Berman,M.S. (2008a) - \textit{A General Relativistic Rotating Evolutionary
Universe, }Astrophysics and Space Science, \textbf{314}, 319-321.

Berman,M.S. (2008b) - \textit{A General Relativistic Rotating Evolutionary
Universe - Part II, }Astrophysics and Space Science, to appear.

Berman,M.S.; Marinho,R.M. (2001) - Astrophysics and Space Science,
\textbf{278}, 367.

Berman,M.S.; Som, M.M. (1993) - Astrophysics and Space Science, \textbf{207}, 105.

Brans, C.; Dicke, R.H. (1961) - Physical Review, \textbf{124}, 925.

\noindent\bigskip\noindent\ \ \ \ \ \ Chen,W.;Wu,Y.- S. (1990) - Physical
Review \textbf{D41},695.

Cooperstock, F.I.; Israelit, M. (1995) - Foundations of Physics, \textbf{25}, 631.

D'Inverno, R. (1992) - \textit{Introducing Einstein's Relativity, }Clarendon
Press, Oxford.

Feynman, R. (1962-3) - \textit{Lectures on Gravitation}, Addison-Wesley, N.Y.

Gomide, F.M.(1963) - Nuovo Cimento, \textbf{30}, 672.

Hawking, S.W. (2001) - \textit{The Universe in a Nutshell, }Bantam, N.Y.

Kolb, E.W.; Turner, M.S. (1990) - \textit{The Early Universe}, Addison-Wesley, N.Y.

Raychaudhuri, A.K. (1979) - \textit{Theoretical Cosmology, }OUP, Oxford.

Rosen, N. (1994) - GRG \textbf{26}, 319.

Sabbata, V.; Sivaram, C. (1994) - \textit{Spin and Torsion in Gravitation,}
World Scientific, Singapore.

Sabbata, V.; Gasperini,M.\ (1979) - Lettere Nuovo Cimento \textbf{25}, 489.

Sciama, D.N. (1953) - MNRAS \textbf{113}, 34.

Sears, F.W.; Salinger,G.L. (1975) - \textit{Thermodynamics, Kinetic theory,
and Statistical Thermodynamics, }Addison-Wesley, New York.

Wesson, P.S. (1999) - \textit{Space-Time-Matter}, World Scientific, Singapore.

Wesson, P.S. (2006) - \textit{Five dimensional Physics}, World Scientific,

\end{document}